# Topological mosaic in Moiré superlattices of van der Waals heterobilayers

Qingjun Tong[1], Hongyi Yu[1], Qizhong Zhu[1], Yong Wang[2,1], Xiaodong Xu[3,4], Wang Yao[1*]

[1]Department of Physics and Center of Theoretical and Computational Physics, University of Hong Kong, Hong Kong, China
[2]School of Physics, Nankai University, Tianjin 300071, China
[3]Department of Physics, University of Washington, Seattle, WA 98195, USA
[4]Department of Materials Science and Engineering, University of Washington, Seattle, WA 98195, USA

*Correspondence to: wangyao@hku.hk

**Van der Waals (vdW) heterostructures formed by 2D atomic crystals provide a powerful approach towards designer condensed matter systems[1-15]. Incommensurate heterobilayers with small twisting and/or lattice mismatch lead to the interesting concept of Moiré superlattice[2-7], where the atomic registry is locally indistinguishable from commensurate bilayers but has local-to-local variation over long range. Here we show that such Moiré superlattice can lead to periodic modulation of local topological order in vdW heterobilayers formed by two massive Dirac materials. By tuning the vdW heterojunction from normal to the inverted type-II regime via an interlayer bias, the commensurate heterobilayer can become a topological insulator (TI), depending on the interlayer hybridization controlled by the atomic registry between the vdW layers. This results in mosaic pattern of TI regions and normal insulator (NI) regions in Moiré superlattices, where topologically protected helical modes exist at the TI/NI phase boundaries. By using symmetry based k.p and tight-binding models, we predict that this topological phenomenon can be present in inverted transition metal dichalcogenides heterobilayers. Our work points to a new means of realizing programmable and electrically switchable topological superstructures from 2D arrays of TI nano-dots to 1D arrays of TI nano-stripes.**

2D topological insulators (TI), or quantum spin Hall (QSH) insulators[16-18], feature topologically protected helical edge states with exotic conducting behaviors in the bulk insulating gap. Experimental evidences of such edge states have been extensively explored in the QSH phases of HgTe/CdTe and InAs/GaSb QWs arising from topological band inversion at the heterojunctions[19-26]. Implementation of the helical conducting channels may have profound consequences in quantum electronics and spintronics, but such channels exist only at the crystal edge or interface with other normal insulator (NI) materials. The typical planar layout of integrated circuits implies the need to engineer programmable lateral superstructures comprised of distributed TI edges or TI/NI interfaces. Meanwhile, the on/off switching of the QSH phases and the helical channels is also highly desirable[27]. Such systems may open new opportunities for manipulating topological phenomena and enable new functionalities for integrated electronics and spintronics.

Compared to existing crystalline QSH systems[18-28], vdW heterobilayers formed by stacking 2D atomic crystals offer several unique possibilities. Some 2D crystals such as monolayer transition metal dichalcogenides (TMD) endow electrons the Dirac physics[29,30]. Using these 2D crystals as building blocks points to interesting scenario to investigate band inversion of Dirac fermions[31,32], for example, an inverted type-II heterojunction as first explored in InAs/GaSb QWs[22,23], but now between massive Dirac cones (i.e. upper cone from one layer energetically overlaps and hybridizes with lower cone from the other layer, c.f. Fig. 1). The tunable heterojunction band alignment by an interlayer bias[9-15] can then realize the desired on/off switch of the possible topological band inversion. The most interesting possibility comes from engineering superlattices in incommensurate bilayers of long period Moiré pattern[2-7]. Within a length scale much larger than the lattice constant but small compared to the Moiré period, the atomic registry between the vdW layers has negligible difference from commensurate bilayers, while the Moiré pattern is the long-range variation in the local registry. With such an atomic registry determining the form of interlayer hybridization, the Moiré pattern can realize a superlattice modulation on the local electronic structures and possibly topological structures.

Here we show that engineering Moiré superlattices in heterobilayers of massive Dirac materials can lead to the concept of programmable topological mosaic (i.e. lateral superstructure modulation of local topological phases). The findings are put forward in several steps. First we show that, upon the interlayer hybridization of massive Dirac cones in the inverted type-II regime, a commensurately stacked bilayer can become a TI, depending on the form of interlayer hopping that is controlled by the atomic registry between the layers. This general finding is embodied by the example of inverted TMD heterobilayers, where symmetry dictates multiple TI phases separated by NI phase when the interlayer atomic registry is varied. We then consider *incommensurate* TMD heterobilayers, in which the formation of long-period Moiré pattern determines the variation in the local atomic registry and thus corresponds to a pattern of spatial separation of the TI and NI phases. Strain can tune such topological mosaic from 2D arrays of TI nano-dots to 1D arrays of TI nano-stripes. Lastly we use tight-binding model to numerically demonstrate the topological mosaic and the topologically protected helical modes at the TI/NI phase boundaries in 1D Moiré superlattices formed by a strained TMD monolayer on an unstrained one.

We start by considering *commensurate* bilayers of typical Dirac materials with the 2D hexagonal lattice, where the Dirac cones sit at the K and -K corners of the hexagonal Brillouin zone, denoted by the valley index $\tau = \pm 1$. The Dirac cones from the two layers and their coupling can be described by the minimal k.p model:

$$H_{\tau=1}(\boldsymbol{q}) = \begin{bmatrix} -\Delta/2 + M_u & v_u(q_x - i\epsilon q_y) & t_{cc} & t_{cv} \\ v_u(q_x + i\epsilon q_y) & -\Delta/2 & t_{vc} & t_{vv} \\ t_{cc}^* & t_{vc}^* & \Delta/2 & v_l(q_x - iq_y) \\ t_{cv}^* & t_{vv}^* & v_l(q_x + iq_y) & \Delta/2 - M_l \end{bmatrix} \quad (1)$$

Here $M_u$ ($M_l$) corresponds to the mass and $v_u$ ($v_l$) is Fermi velocity in the upper (lower) layer. $\epsilon = \pm$ denotes two types of valley-alignment of the Dirac cones in the bilayer, determined by the orientation of stacking. In the example of TMD heterobilayers, $\epsilon = -$ corresponds to the H-type stacking (Fig. 2a), while $\epsilon = +$ is for the R-type stacking (Fig. 2b). $t_{ij}$ ($i,j = c, v$) are the interlayer hopping matrix elements between the $q = 0$ (K point) Bloch state in band $i$ of layer $u$ and the one in band $j$ of layer $l$. $\Delta = \Delta_g - U$, where $U$ is the interlayer bias proportional to the perpendicular electric field, and $\Delta_g$ is the bilayer band gap in absence of field (c.f. Supplementary Table S5).

Fig. 1a schematically illustrates the type-II heterojunction between the Dirac cones from the two layers. For $\Delta < 0$, the bilayer is in the *inverted* type-II regime, i.e. the lower cone in layer $u$ and upper cone in layer $l$ energetically overlap. The cones from the two layers can then hybridize through the four possible interlayer hopping channels as given in Eq. (1). We focus on the neighborhood of the critical point $\Delta = 0$. In such case, $M_u, M_l \gg \Delta$ and the 4x4 Hamiltonian in Eq. (1) can be projected to a 2x2 one,

$$H(\boldsymbol{q}) \cong \left(\frac{\Delta}{2} + \frac{B}{2}q^2\right)\hat{\sigma}_z + \frac{D}{2}q^2 + \left[\hat{\sigma}_+ \left(t_{vc}^* + \frac{v_l}{M_l}t_{vv}^* q_- - \frac{v_u}{M_u}t_{cc}^* q_{-\epsilon} + \frac{v_l v_u}{M_l M_u}t_{cv}^* q_{-\epsilon} q_-\right) + h.c.\right], \quad (2)$$

where $B = \left(\frac{v_u^2}{M_u} + \frac{v_l^2}{M_l}\right)$, $D = \frac{v_u^2}{M_u} - \frac{v_l^2}{M_l}$, $q \equiv |\boldsymbol{q}|$, $q_\pm \equiv q_x \pm i q_y$ and $\hat{\sigma}_\pm \equiv \hat{\sigma}_x \pm i\hat{\sigma}_y$. The Pauli matrices $\hat{\boldsymbol{\sigma}}$ are spanned by the $q = 0$ conduction state in layer $l$ and valence state in layer $u$. For the interlayer hopping effect, we only retain up to the leading order term by each hopping channel.

As $\Delta$ is varied from positive to negative by the interlayer bias, the Hamiltonian in Eq. (2) can have a gap-closing topological phase transition depending on the dominance of the interlayer hopping terms. The $t_{vc}$ term has a $q$-independent coupling form so its dominance in the interlayer hybridization leads to the avoided band crossing (gap never closes) where the Hall conductance in the gap remains zero (Fig. 1b). For the R-type stacking ($\epsilon = +$), the $t_{cc}$ and $t_{vv}$ terms in Eq. (2) both have a $q$-linear coupling with chirality index of 1, so their dominance leads to a topological band inversion where the Hall conductance changes by a quantized value $-e^2/h$ at the gap closing point (Fig. 1c). The $t_{cv}$ term has a $q$-quadratic coupling with chirality index of 2, so its dominance leads to topological band inversion where the Hall conductance changes by $-2e^2/h$. For the H-type stacking ($\epsilon = -$), both $t_{vc}$ and $t_{cv}$ terms lead to the trivial avoided crossing, while $t_{cc}$ ($t_{vv}$) term leads to a topological band inversion where the Hall conductance changes by $e^2/h$ ($-e^2/h$).

The relative strength among the different interlayer hopping matrix elements $t_{ij}$ are determined by the atomic configuration including the orbital composition of the Dirac cones and the lateral registry between the vdW layers. Varying the latter by an interlayer translation therefore becomes a control knob to tune between the different topological phases in the inverted regime. This is demonstrated below using the realistic example of TMD heterobilayers. The valence band edge of TMD monolayer has spin up (down) state only at the K (-K) valley because of the giant spin splitting. So only the spin up (down) massive Dirac cones are relevant at K (-K) valley. First principle calculations and experiments have shown that certain TMD heterobilayers can be tuned into the inverted regime by a moderately strong interlayer bias [33,34] (c.f. Supplementary Table S5).

We first analyze high symmetry stacking configurations of the H-type (i.e. two layers have opposite orientations). If either the metal (M) or chalcogen (X) sites of layer $u$ are vertically aligned with M or X sites of layer $l$, the C$_3$ rotational symmetry is retained. For the three high symmetry H-configurations (Fig. 2a), the rotational symmetry of the Bloch states at $\pm K$ dictates that certain interlayer hopping channels must vanish, as listed in Table 1. It is then clear from Eq. (2) that inverted TMD heterobilayers of $H_M^M$ stacking is a topological insulator with a quantized spin Hall conductance $\sigma_s = -1$ (in unit of $e^2/h$) in the hybridization gap. TMD heterobilayer of $H_X^M$ stacking is also a topological insulator, but with an opposite spin Hall conductance ($\sigma_s = 1$), while $H_X^X$ heterobilayer is a normal insulator ($\sigma_s = 0$).

All stacking configurations of the same orientation can span a phase space parameterized by the interlayer translation $r_0$, defined as the lateral displacement between two metal sites from the two layers respectively, which takes values within a unit cell. With the configurations $H_M^M$, $H_X^M$ and $H_X^X$ being three separated points in this phase space (Fig. 2a), the conclusion that they have distinct QSH conductance from the above symmetry analysis means the inverted TMD bilayer undergoes topological phase transitions as $r_0$ is varied.

With these symmetry dictated qualitative features of the topological phase diagram, the remaining details are determined by the dependence of interlayer hopping $t_{ij}$ on the interlayer translation $r_0$. In TMD bilayers, the conduction states near $q = 0$ are predominantly from the M $d$-orbital with magnetic quantum number $m = 0$, while the valence states at valley $\tau$ are from the M $d$-orbital with $m = 2\tau$. The $r_0$ dependence of $t_{ij}$ can be approximated as[8]

$$t_{cc}(r_0) \cong T_0^0 \left( e^{iK \cdot r_0} + e^{i\hat{C}_3 K \cdot r_0} + e^{i\hat{C}_3^2 K \cdot r_0} \right), \quad (3)$$

$$t_{vc}(r_0) \cong T_0^{2\epsilon} \left( e^{iK \cdot r_0} + e^{i(\hat{C}_3 K \cdot r_0 + 2\epsilon\pi/3)} + e^{i(\hat{C}_3^2 K \cdot r_0 - 2\epsilon\pi/3)} \right),$$

$$t_{cv}(r_0) \cong T_2^0 \left( e^{iK \cdot r_0} + e^{i(\hat{C}_3 K \cdot r_0 - 2\pi/3)} + e^{i(\hat{C}_3^2 K \cdot r_0 + 2\pi/3)} \right),$$

$$t_{vv}(r_0) \cong T_2^{2\epsilon} \left( e^{iK \cdot r_0} + e^{i(\hat{C}_3 K \cdot r_0 + (1-\epsilon)\pi/3)} + e^{i(\hat{C}_3^2 K \cdot r_0 - (1-\epsilon)\pi/3)} \right).$$

where $K$, $\hat{C}_3 K$ and $\hat{C}_3^2 K$ are respectively the wavevectors for the three K corners of the first Brillouin Zone. $T_m^{m'}$ denotes the Fourier component at $K$ of the interlayer hopping integral between a $d_{m'}$ orbital in layer $u$ and a $d_m$ orbital in layer $l$. Note that we derived Eq. (3) under the two-center approximation and have dropped Umklapp terms that are generally expected to be weak (see Supplementary text Ib). As shown later, quantitative details subject to this approximation are nonessential to the conclusions, as long as $t_{ij}(\boldsymbol{r}_0)$ observes the C$_3$ symmetry dictated behavior including those listed in Table 1.

Fig. 2a shows the phase diagram for the inverted H-type TMD bilayers. In the calculations, $T_0^0 = 6.7$meV, $T_0^{-2} = T_2^0 = 3.3$meV, and $T_2^{-2} = 10$meV which are estimated from the first principle band structures of TMD homobilayers[35,36] (see Supplementary text Ic). The color map plots the hybridization gap $\delta$ (c.f. Fig. 1b-c), which is taken as negative (positive) if the Hall conductance is quantized (zero) in the gap. There are two isolated (red) regions of negative $\delta$ (i.e. TI phases) centered respectively at the $H_M^M$ and $H_X^M$ points. The rest (blue) is region of positive $\delta$ (i.e. NI). Fig. 2c shows the electrically controlled topological phase transition for the stacking configurations along the dashed horizontal line in Fig. 2a. The TI phases start to appear at $\Delta = 0$ at the high symmetry $H_M^M$ and $H_X^M$ points, while the size of the TI phase regions grows with the magnitude of the inverted gap ($|\Delta|$). The insets are the hybridized Dirac cones calculated from Eq. (1) at various values of interlayer translation and $\Delta$, which clearly illustrate the avoided band crossing in $H_X^X$ TMD bilayer and the topological band inversion in $H_M^M$ TMD bilayer as $\Delta$ changes, as well as the asymmetric band touching upon the topological phase transition in TMD bilayers of low symmetry stacking.

For the R-type stacking (two layers have same orientations), there is similar topological phase diagram. For the high symmetry $R_M^M$ stacking configuration, $t_{cc}$ and $t_{vv}$ are the only symmetry allowed interlayer hopping channels in Eq. (1), which dictate the inverted TMD heterobilayer to be a TI with $\sigma_s = -1$. $R_X^M$ TMD bilayer is always a NI. Fig. 2b shows the calculated phase diagram based on Eq. (1) and (3), where we find a TI phase (red) region and a NI phase (blue) region centered respectively at $R_M^M$ and $R_X^M$ points. Eq. (1) also predicts that the $R_M^X$ TMD bilayer is a TI with $\sigma_s = -2$ in the inverted regime. However, compared to other TI phases, the topological band inversion at $R_M^X$ is through a weak higher order effect (c.f. Eq. (2)). When other interlayer hopping channels through bands beyond the massive Dirac cones are considered, $R_M^X$ TMD bilayer in the inverted regime can become a TI with a different $\sigma_s$ value (see Supplementary text IIb).

To demonstrate the QSH edge states in the TI phases, we turn to tight-binding (TB) calculations. Our bilayer TB model is generalized from the nearest-neighbor three-orbital TB model for the monolayer[37], by adding interlayer hopping that observes the rotational symmetry of the *d*-orbitals (see Supplementary text IIa). For simplicity, the two-center interlayer integrals used are short-range ones allowing hopping to three neighboring metal sites in the opposite layer only. With this simplification, Umklapp terms have pronounced contribution to the band-edge hopping matrix elements $t_{ij}$, which results in different $r_0$-dependences from Eq. (3) (c.f. Supplementary Fig. S2). Nevertheless, comparing the phase diagrams from the k.p (Fig. 2a) and from this TB calculations (Fig. 3i), all important features are in excellent agreement, including the locations and the areas of the TI phase regions and the values of QSH conductance. This shows that the topological phase diagram is determined by the symmetry in the TMD bilayer, rather than quantitative details of the models.

Using the TB model, we can examine electronic structures at the edges or interfaces with normal insulating regimes. Fig. 3a-h shows the TB calculations for several H-type TMD bilayers in the inverted regime, demonstrating the evolution of the bands and the emergence of the QSH edge states in the hybridization gap as the interlayer translation is varied across the TI/NI phase boundaries.

Now we turn to incommensurate TMD heterobilayers where lattice mismatch and twisting lead to Moiré pattern. The Moiré supercell period can reach ~ 10 - 100 nm in TMD heterobilayers of small twisting angles, depending on the choices of the compounds. In such long-period Moiré, within a length scale smaller than the Moiré period (*b*) but much larger than the lattice constant (*a*), the local atomic registry has negligible difference from commensurate TMD bilayer of a certain interlayer translation $r_0$ (c.f. insets in Fig. 4a). Studies of twisted bilayer graphene and graphene/hBN have shown that[7,38,39], in the limit of Moiré period ≫ lattice constant, the local electronic structure at any location $R$ is well described by the bands of commensurate bilayer with the corresponding interlayer translation $r_0(R)$, which, in the present context of TMD bilayer, are topologically nontrivial (trivial) if the local hybridization gap $\delta(r_0)$ is negative (positive). For example, for the three magnified local regions in Fig. 4a, $R_1$ and $R_2$ have the electronic structure of the TI phase while $R_3$ has that of the NI phase. Therefore topological phase separations are expected in the Moiré superlattice. The color maps in Fig. 4 are plots of $\delta(\boldsymbol{r}_0(\boldsymbol{R}))$, the position dependent hybridization gap from this local approximation, in various Moiré patterns which form superlattice modulations of the local topological orders.

When the Moiré is formed between two undistorted hexagonal lattices, the phase separation in a Moiré supercell is isomorphic to the topological phase diagram of commensurate TMD bilayer parameterized by $\boldsymbol{r}_0$ (Fig. 2 and 3). Fig. 4a-b shows the topological phase separation in such Moiré, realizing periodic arrays of TI nano-dots, where helical modes circulating along the phase boundaries are expected inside the bulk hybridization gap. The variation of the local atomic registry in the Moiré, as described by the function $\boldsymbol{r}_0(\boldsymbol{R})$, can be tuned by the twisting angle as well as strain. This further allows the tuning of the pattern of topological phase separations, for example, from 2D arrays of TI nano-dots in Fig. 4b to 1D arrays of TI nano-stripes in Fig. 4d.

We numerically demonstrate the topological mosaic and the topologically protected helical modes at the TI/NI phase boundaries using TB calculations. Here we consider H-type stacking of two TMD layers with identical lattice period in $y$ (zigzag) direction and a lattice mismatch $\eta$ in $x$ (armchair) direction. In such 1D Moiré superlattices, the local approximation predicts arrays of TI nano-stripes (c.f. Fig. 4d). The two-center integrals used in the TB calculations here are the same as those used for Fig. 3, but generate locally different interlayer hopping due to the local-to-local variation of atomic registry. Compared to the calculations in Fig. 3, we have used larger interlayer hopping integrals $T_0^0 = 20$meV, $T_0^{-2} = T_2^0 = 10$meV, and $T_2^{-2} = 30$meV (achievable at a smaller interlayer distance, c.f. Supplementary text I), to allow a larger hybridization gap to examine the behaviors of the in-gap modes. The strain effect on intralayer hopping due to change in the atomic positions is also incorporated (see Supplementary text III).

Fig. 5 shows the calculated energy spectra as functions of wavevector $k$ in the translational invariant $y$-direction, for Moiré superlattices with various lattice mismatch $\eta$ in $x$ direction. For small $\eta$ value, gapless helical modes are indeed found in the gap of bulk spectra (c.f. Fig. 5a-b). For the helical modes at an in-gap energy $E_F$, their distribution in the $x$ direction are plotted, together with the spatial color map of local hybridization gap $\delta(\boldsymbol{r}_0(\boldsymbol{R}))$. These modes, labeled by their wavevectors in $y$-direction ($k_1, k_2, k_3, k_4$), are all localized at the $x$-locations corresponding to the zeros of $\delta(\boldsymbol{r}_0(\boldsymbol{R}))$. The gapless modes at valley -K are all spin down states, and their time-reversal counterparts are at valley K. Their chirality is consistent with the quantum spin Hall conductance associated with $\delta(\boldsymbol{r}_0(\boldsymbol{R}))$ (c.f. Fig. 2c and Table 1). This confirms the topological phase separations in these long period Moiré, where locations of the TI/NI boundaries and the helical modes are well predicted by $\delta(\boldsymbol{r}_0(\boldsymbol{R}))$ from the local approximation.

Fig. 5c shows the evolution of the spectra with the increase of $\eta$. The change in the band alignment arises from the strain effect on the intralayer hopping. When the Moiré period (and hence the width of the TI nano-stripes) are not large enough, the helical modes on different sides of a nano-stripe can couple, which leads to a finite gap at their crossing in the energy-momentum space, as shown in Fig. 5c. Through such finite size effect, the helical modes at the TI/NI phase boundaries in long-period Moiré will eventually evolve into minibands in short-period Moiré.

**Reference**

1   Geim, A. K. & Grigorieva, I. V. Van der Waals heterostructures. *Nature* **499**, 419-425, doi:10.1038/nature12385 (2013).
2   Ponomarenko, L. A. *et al.* Cloning of Dirac fermions in graphene superlattices. *Nature* **497**, 594-597, doi:10.1038/nature12187 (2013).
3   Dean, C. R. *et al.* Hofstadter's butterfly and the fractal quantum Hall effect in moiré superlattices. *Nature* **497**, 598-602, doi:10.1038/nature12186 (2013).
4   Hunt, B. *et al.* Massive Dirac Fermions and Hofstadter Butterfly in a van der Waals Heterostructure. *Science* **340**, 1427-1430, doi:10.1126/science.1237240 (2013).
5   Gorbachev, R. V. *et al.* Detecting topological currents in graphene superlattices. *Science* **346**, 448-451, doi:10.1126/science.1254966 (2014).
6   Song, J. C. W., Samutpraphoot, P. & Levitov, L. S. Topological Bloch bands in graphene superlattices. *Proc. Natl. Acad. Sci. USA* **112**, 10879-10883, doi:10.1073/pnas.1424760112 (2015).
7   Jung, J., Raoux, A., Qiao, Z. & MacDonald, A. H. Ab initio theory of moiré superlattice bands in layered two-dimensional materials. *Phys. Rev. B* **89**, 205414, doi:10.1103/PhysRevB.89.205414 (2014).
8   Yu, H., Wang, Y., Tong, Q., Xu, X. & Yao, W. Anomalous light cones and valley optical selection rules of interlayer excitons in twisted heterobilayers. *Phys. Rev. Lett.* **115**, 187002, doi:10.1103/PhysRevLett.115.187002 (2015).
9   Rivera, P. *et al.* Valley-polarized exciton dynamics in a 2D semiconductor heterostructure. *Science* **351**, 688-691, doi:10.1126/science.aac7820 (2016).
10  Fang, H. *et al.* Strong interlayer coupling in van der Waals heterostructures built from single-layer chalcogenides. *Proc. Natl. Acad. Sci. USA* **111**, 6198-6202, doi:10.1073/pnas.1405435111 (2014).
11  Chiu, M.-H. *et al.* Spectroscopic Signatures for Interlayer Coupling in MoS2-WSe2 van der Waals Stacking. *ACS Nano* **8**, 9649-9656, doi:10.1021/nn504229z (2014).

12    Lee, C.-H. *et al.* Atomically thin p-n junctions with van der Waals heterointerfaces. *Nature Nanotech.* **9**, 676-681, doi:10.1038/nnano.2014.150 (2014).
13    Furchi, M. M., Pospischil, A., Libisch, F., Burgdörfer, J. & Mueller, T. Photovoltaic Effect in an Electrically Tunable van der Waals Heterojunction. *Nano Lett.* **14**, 4785-4791, doi:10.1021/nl501962c (2014).
14    Cheng, R. *et al.* Electroluminescence and Photocurrent Generation from Atomically Sharp WSe2/MoS2 Heterojunction p-n Diodes. *Nano Lett.* **14**, 5590-5597, doi:10.1021/nl502075n (2014).
15    Hong, X. *et al.* Ultrafast charge transfer in atomically thin MoS2/WS2 heterostructures. *Nature Nanotech.* **9**, 682-686, doi:10.1038/nnano.2014.167 (2014).
16    Qi, X.-L. & Zhang, S.-C. Topological insulators and superconductors. *Rev. Mod. Phys.* **83**, 1057, doi:10.1103/RevModPhys.83.1057 (2011).
17    Hasan, M. Z. & Kane, C. L. Colloquium: Topological insulators. *Rev. Mod. Phys.* **82**, 3045, doi:10.1103/RevModPhys.82.3045 (2010).
18    Kane, C. L. & Mele, E. J. Quantum Spin Hall Effect in Graphene. *Phys. Rev. Lett.* **95**, 226801, doi:10.1103/PhysRevLett.95.226801 (2005).
19    Bernevig, B. A., Hughes, T. L. & Zhang, S.-C. Quantum Spin Hall Effect and Topological Phase Transition in HgTe Quantum Wells. *Science* **314**, 1757-1761, doi:10.1126/science.1133734 (2006).
20    König, M. *et al.* Quantum Spin Hall Insulator State in HgTe Quantum Wells. *Science* **318**, 766-770, doi:10.1126/science.1148047 (2007).
21    Ma, E. Y. *et al.* Unexpected edge conduction in mercury telluride quantum wells under broken time-reversal symmetry. *Nat. Commun.* **6**, 7252, doi:10.1038/ncomms8252 (2015).
22    Liu, C., Hughes, T. L., Qi, X.-L., Wang, K. & Zhang, S.-C. Quantum Spin Hall Effect in Inverted Type-II Semiconductors. *Phys. Rev. Lett.* **100**, 236601, doi:10.1103/PhysRevLett.100.236601 (2008).
23    Knez, I., Du, R.-R. & Sullivan, G. Evidence for Helical Edge Modes in Inverted InAs/GaSb Quantum Wells. *Phys. Rev. Lett.* **107**, 136603, doi:10.1103/PhysRevLett.107.136603 (2011).
24    Karalic, M. *et al.* Experimental Evidence for the Topological Insulator Phase in InAs/GaSb Coupled Quantum Wells. *arXiv:1606.03627* (2016).
25    Nichele, F. *et al.* Edge Transport in the Trivial Phase of InAs/GaSb. *arXiv:1511.01728* (2016).
26    Suzuki, K., Harada, Y., Onomitsu, K. & Muraki, K. Edge channel transport in the InAs/GaSb topological insulating phase. *Phys. Rev. B* **87**, 235311, doi:10.1103/PhysRevB.87.235311 (2013).
27    Qian, X., Liu, J., Fu, L. & Li, J. Quantum spin Hall effect in two-dimensional transition metal dichalcogenides. *Science* **346**, 1344-1347, doi:10.1126/science.1256815 (2014).
28    Abanin, D. A., Lee, P. A. & Levitov, L. S. Spin-Filtered Edge States and Quantum Hall Effect in Graphene. *Phys. Rev. Lett.* **96**, 176803, doi:10.1103/PhysRevLett.96.176803 (2006).
29    Xiao, D., Liu, G.-B., Feng, W., Xu, X. & Yao, W. Coupled Spin and Valley Physics in Monolayers of MoS 2 and Other Group-VI Dichalcogenides. *Phys. Rev. Lett.* **108**, 196802, doi:10.1103/PhysRevLett.108.196802 (2012).

30  Mak, K. F., McGill, K. L., Park, J. & McEuen, P. L. The valley Hall effect in MoS2 transistors. *Science* **344**, 1489-1492, doi:10.1126/science.1250140 (2014).
31  Sie, E. J. *et al.* Valley-selective optical Stark effect in monolayer WS2. *Nature Mater.* **14**, 290-294, doi:10.1038/nmat4156 (2014).
32  Claassen, M., Jia, C., Moritz, B. & Devereaux, T. P. All-Optical Materials Design of Chiral Edge Modes in Transition-Metal Dichalcogenides. *arXiv:1603.04457* (2016).
33  Gong, C. *et al.* Band alignment of two-dimensional transition metal dichalcogenides: Application in tunnel field effect transistors. *Appl. Phys. Lett.* **103**, 053513, doi:10.1063/1.4817409 (2013).
34  Chiu, M.-H. *et al.* Determination of band alignment in the single-layer MoS2/WSe2 heterojunction. *Nat. Commun.* **6**, 7666, doi:10.1038/ncomms8666 (2015).
35  Liu, K. *et al.* Evolution of interlayer coupling in twisted molybdenum disulfide bilayers. *Nat. Commun.* **5**, 4966, doi:10.1038/ncomms5966 (2014).
36  Gong, Z. *et al.* Magnetoelectric effects and valley-controlled spin quantum gates in transition metal dichalcogenide bilayers. *Nat. Commun.* **4**, 2053, doi:10.1038/ncomms3053 (2013).
37  Liu, G.-B., Shan, W.-Y., Yao, Y., Yao, W. & Xiao, D. Three-band tight-binding model for monolayers of group-VIB transition metal dichalcogenides. *Phys. Rev. B* **88**, 085433, doi:10.1103/PhysRevB.88.085433 (2013).
38  Santos, J. M. B. L. d., Peres, N. M. R. & Neto, A. H. C. Graphene Bilayer with a Twist: Electronic Structure. *Phys. Rev. Lett.* **99**, 256802, doi:10.1103/PhysRevLett.99.256802 (2007).
39  Mele, E. J. Band symmetries and singularities in twisted multilayer graphene. *Phys. Rev. B* **84**, 235439, doi:10.1103/PhysRevB.84.235439 (2011).

**Acknowledgments:** The work is mainly supported by the Croucher Foundation (Croucher Innovation Award), the Research Grants Council and University Grants Committee of Hong Kong (HKU17312916, AoE/P-04/08), and the University of Hong Kong (ORA). Y.W. is partly supported by NSFC with Grant No. 11604162 and Grant No.61674083. X.X. is supported by Department of Energy, Basic Energy Sciences, Materials Sciences and Engineering Division (DE-SC0008145 and SC0012509), and the Cottrell Scholar Award.

**Author contributions:** W.Y. conceived and designed the research. Q.T. and H.Y. performed the calculations. Q.T., H.Y. and W.Y. analyzed the results with input from Q.Z. and X.X.. Y.W. provided support with first-principles calculation. W.Y., Q.T., H.Y. and X.X. prepared the manuscript.

**Data availability:** The data that support the plots within this paper and other findings of this study are available from the corresponding author upon reasonable request.

# Tables

|  | $t_{cc}$ | $t_{vv}$ | $t_{cv}$ | $t_{vc}$ | $\sigma_s$ |
|---|---|---|---|---|---|
| $R^M_M$ | $3T^0_0$ | $3T^2_2$ | 0 | 0 | $-1$ |
| $R^M_X$ | 0 | 0 | 0 | $3T^2_0$ | 0 |
| $R^X_M$ | 0 | 0 | $3T^0_2$ | 0 | * |
| $H^M_X$ | 0 | $3T^{-2}_2$ | 0 | 0 | 1 |
| $H^M_M$ | $3T^0_0$ | 0 | 0 | 0 | $-1$ |
| $H^X_X$ | 0 | 0 | $3T^0_2$ | $3T^{-2}_0$ | 0 |

**Table 1. Interlayer hopping channels and QSH conductance in the inverted band of commensurate TMD bilayers with 3-fold rotational symmetry.** $R^\mu_\nu$ ($H^\mu_\nu$) denotes a R-type (H-type) stacking with $\mu$-sites of layer $u$ vertically aligned with $\nu$-sites of layer $l$, where $\mu, \nu = M$ or $X$. See Figure 2. $t_{ij}$ ($i,j = $ c, v) are the interlayer hopping matrix elements between the Bloch state at K point in band $i$ of layer $u$ and the one in band $j$ of layer $l$. $\sigma_s$ is the spin Hall conductance, in unit of $e^2/h$, in the hybridization gap of the inverted heterobilayer. *We note that the massive Dirac model in Eq. (1) leads to $\sigma_s = -2$ through a high order effect, but the value of $\sigma_s$ can be corrected in the presence of other bands (see supplementary information).

Figures:

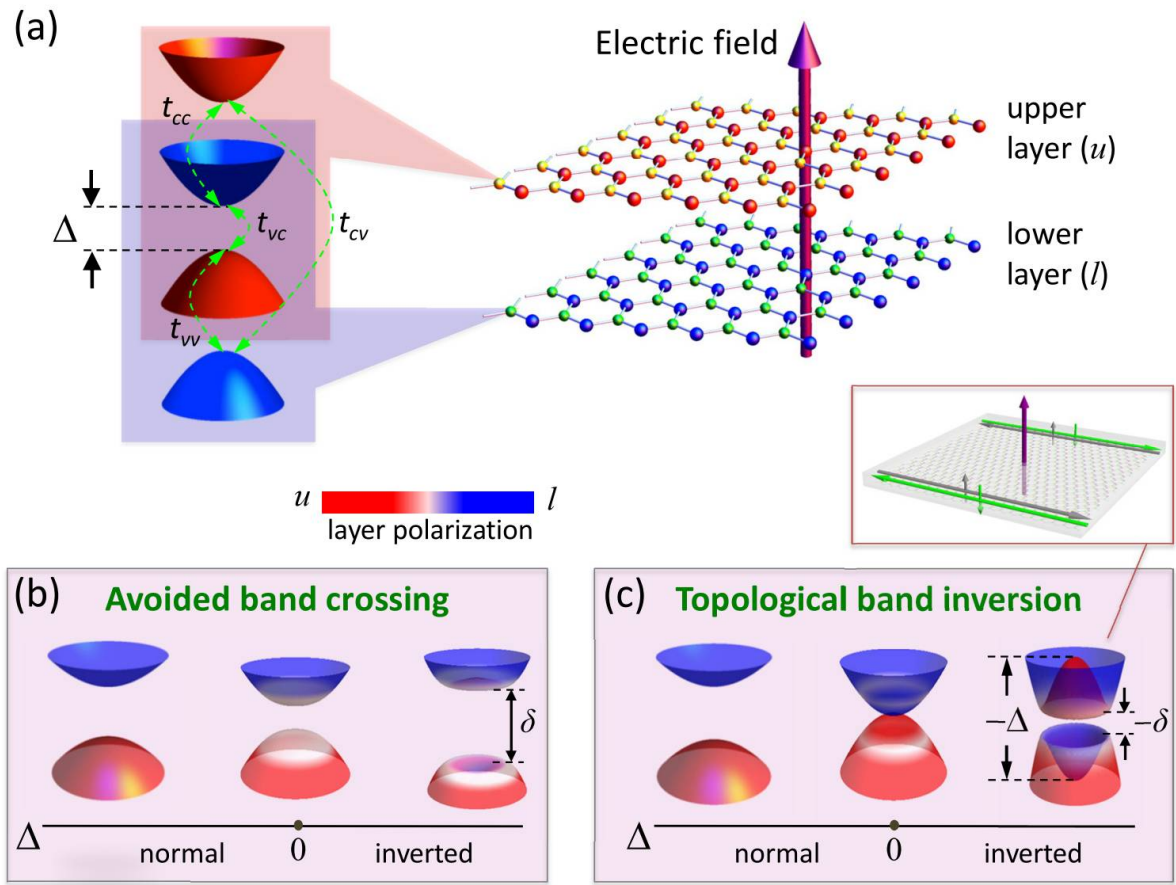

**Figure 1 Electrically controlled band inversion in heterobilayer of massive Dirac materials**. **a**, In type-II band alignment, the heterobilayer band gap $\Delta$, formed between the upper Dirac cone (blue) in layer $l$ and the lower cone (red) in layer $u$, is tunable by a perpendicular electric field. **b** and **c**, When the band alignment is tuned from normal ($\Delta > 0$) to inverted regime ($\Delta < 0$), depending on the relative strength among the interlayer hopping channels (curved arrows in **a**), the Dirac cones from the two layers hybridize either through avoided band crossing OR topological band inversion. In the latter case, the electric field can switch on the topological insulating phase where helical edge modes (c.f. inset) then appear in the hybridization gap $\delta$.

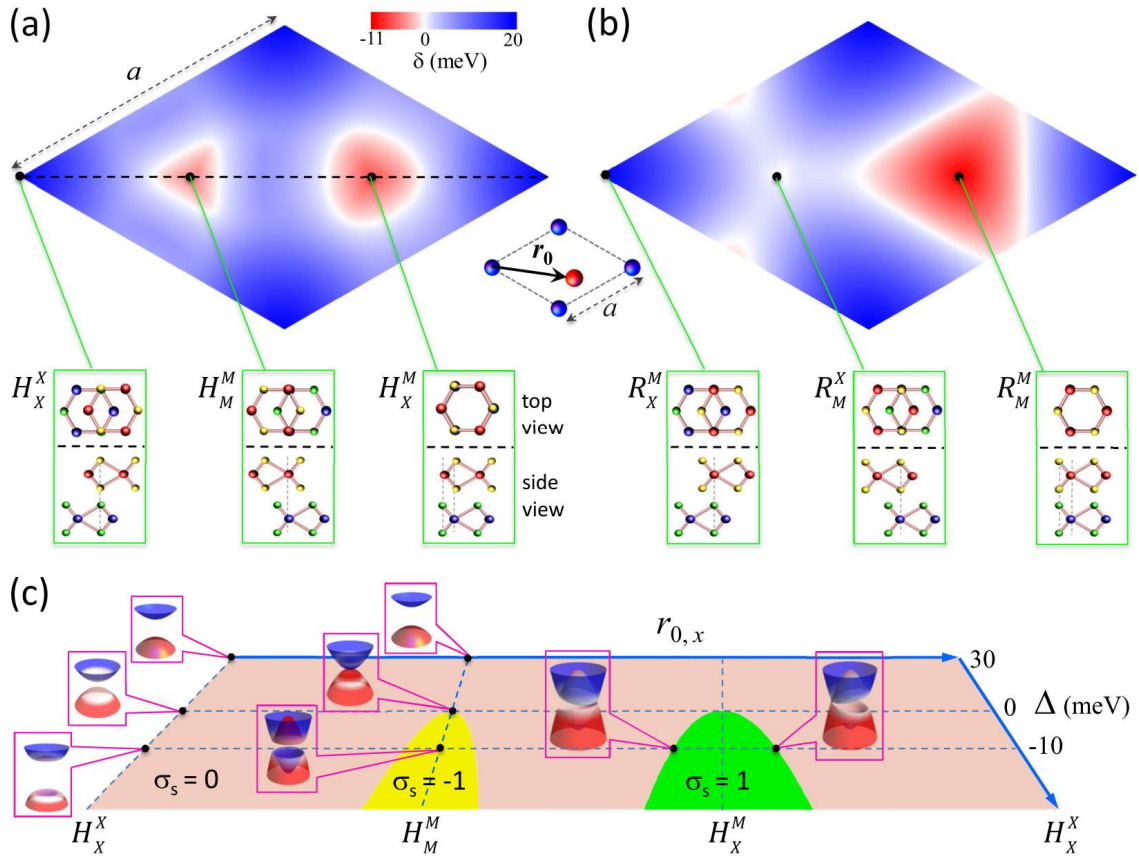

**Figure 2 Topological phase diagrams for commensurate TMD bilayers**. **a**, Hybridization gap $\delta$ as a function of interlayer translation $r_0$ in H-type bilayer (c.f. insets) of inverted type-II alignment ($\Delta = -40$ meV). The bilayer is a topological (normal) insulator in the red (blue) region of the $r_0$ space. Calculation is based on the k.p model (Eq. 1 & 3 in text). **b**, R-type stacking, also at $\Delta = -40$ meV. **c**, Phase diagram parameterized by both $\Delta$ and $r_0$ for the H-type stacking, where $r_0$ is restricted on the dashed line in **a**. Yellow and green regions are two topological insulating phases, with distinct quantum spin Hall conductance $\sigma_s$ in the hybridization gap. At a few representative phase space points, the hybridized Dirac cones at valley -K are shown in the insets.

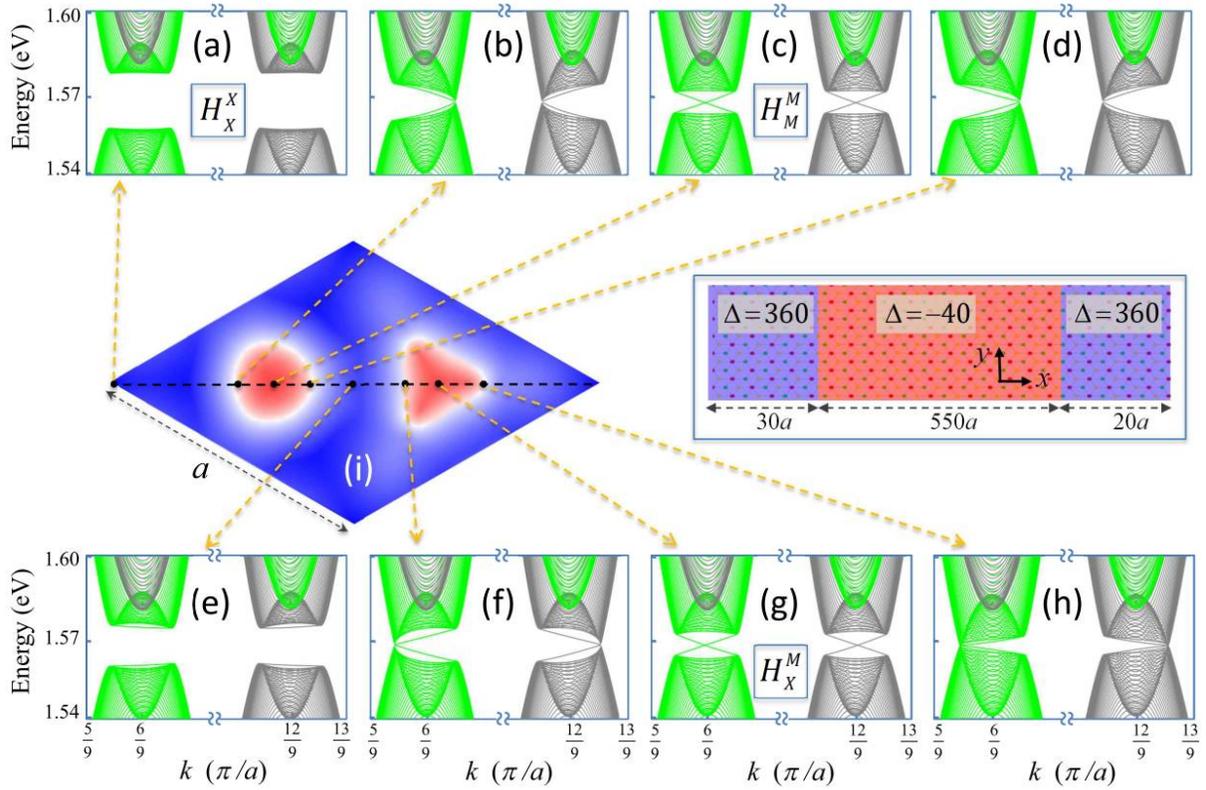

**Figure 3 Bulk-edge correspondence in the topological phase diagram**. **a-h**, Bulk bands and corresponding edge states dispersions in inverted H-type TMD bilayers with various interlayer translation (black dots in **i**). Gray and green colors denote respectively the spin up and down bands. **i**, Bulk hybridization gap $\delta$ from the tight-binding calculation. The sign of $\delta$ is determined by whether there exist gapless edge states. The inset illustrates the setup of the tight-binding calculation. The inverted region with $\Delta = -40$ meV is sandwiched between normal regions with a large gap $\Delta = 360$ meV. The spectra in **a-h** are within the gap of the normal region. The gapless helical modes in the hybridization gap in **c** and **g** are localized at the inverted/normal interfaces.

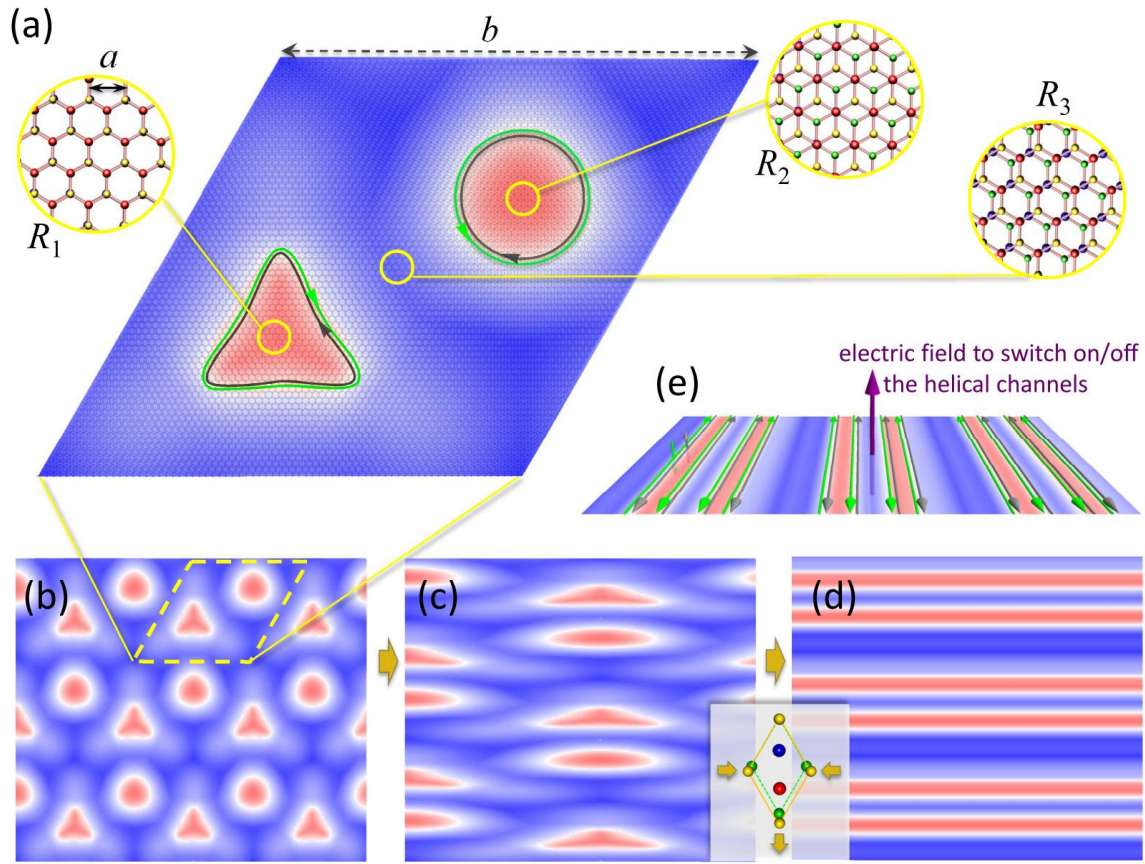

**Figure 4 Topological mosaic in long-period Moiré pattern**. **a**, At three locations (yellow circles) in a Moiré super cell, in a length scale ≫ $a$ (lattice constant) and ≪ $b$ (Moiré period), local atomic registries are shown which have negligible difference from commensurate TMD bilayers with a corresponding interlayer translation $\boldsymbol{r}_0$. Different locals are thus in either topological (red) or normal (blue) insulating phase, depending on the local $\boldsymbol{r}_0$ value. Gray and green colors denote respectively the spin up and down helical modes expected at the TI/NI phase boundaries. **b-d**, Evolution of the topological phase separation patterns, when the lattice mismatch in an armchair (zigzag) direction is increased (reduced) by an uniaxial tensile strain on one layer (c.f. inset). **e**, The electrically switchable dense array of protected helical channels can be exploited for field effect transistor.

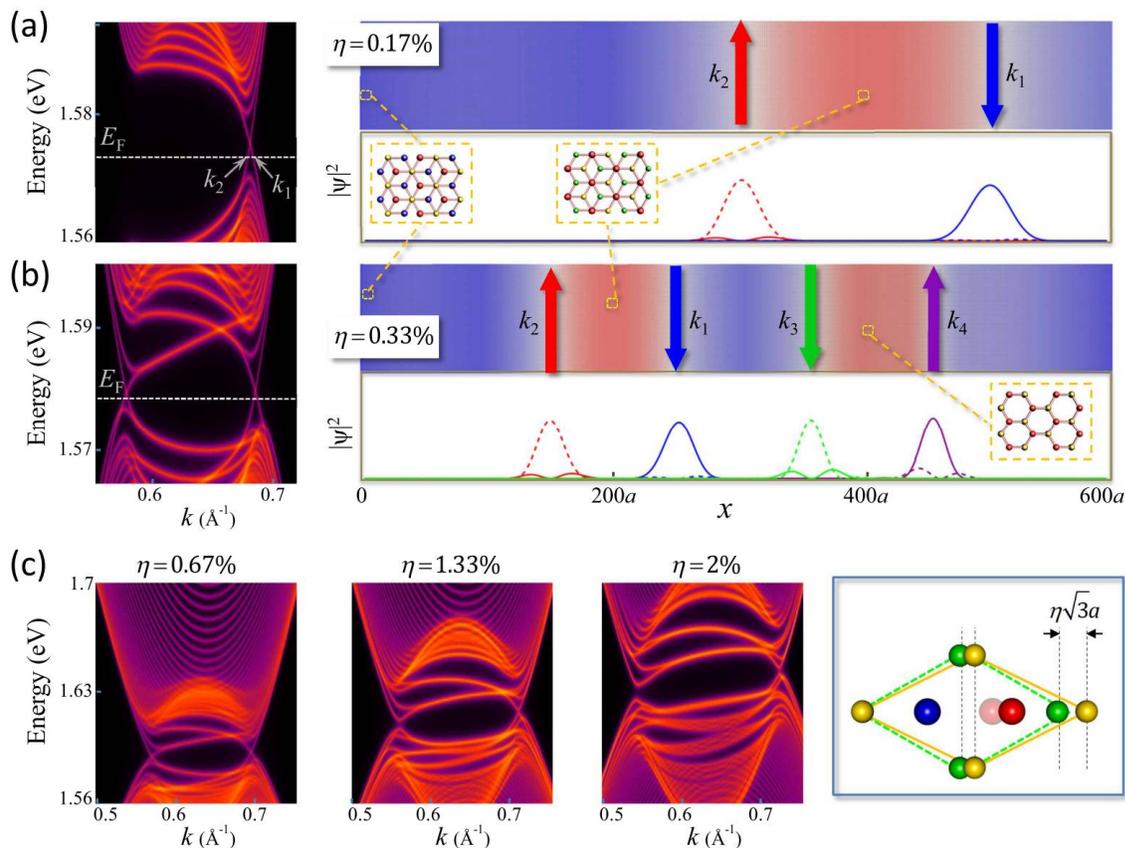

**Figure 5 Topologically protected helical modes at the boundaries of TI nano-stripes in 1D Moiré superlattices**. The two layers have identical lattice period in $y$ (zigzag) direction, while in $x$ (armchair) direction they have a lattice mismatch $\eta$, as defined in the inset. **a**, Left panel plots energy spectra as a function of momentum in $y$-direction, for spin down states at valley -K. Right panel shows density distribution at layer $u$ (solid curves) and layer $l$ (dashed curves) of two in-gap states at energy $E_F$ with momenta $k_1$ (blue) and $k_2$ (red) respectively, while the color map atop plots $\delta(\boldsymbol{r}_0(\boldsymbol{R}))$ from the local approximation (see text). **b**, Same plot at a larger $\eta$. $E_F$ (horizontal line) crosses the in-gap dispersions from right to left at $k_1$, $k_2$, $k_3$ and $k_4$. **c**, With the increase of $\eta$, the in-gap modes become gapped at their crossing point in the energy-momentum space, due to the finite size of the TI stripes.